\documentstyle[prb,aps,multicol]{revtex}
  \renewcommand{\narrowtext}{\begin{multicols}{2} \global\columnwidth20.5pc}
  \renewcommand{\widetext}{\end{multicols} \global\columnwidth42.5pc}

 \input{psfig}

\begin{document}

\draft



\preprint{CondMat}

\title{AC-Hopping Conductance of Self-Organized Ge/Si Quantum Dot Arrays}

\author{Irina L. Drichko, Andrey M. Diakonov, Veniamin I. Kozub, Ivan Yu. Smirnov}

\address{A. F. Ioffe  Physico-Technical Institute of Russian
  Academy of Sciences, 194021
  St. Petersburg, Russia}
\author{Yuri M. Galperin}
\address{Department of Physics, University of Oslo, PO Box
1048
  Blindern, 0316 Oslo, Norway}
\address{A. F. Ioffe  Physico-Technical Institute of Russian
  Academy of Sciences, 194021
  St. Petersburg, Russia}

\author{Andrew I. Yakimov, Alexander I. Nikiforov}
\address{Institute of Semiconductor Physics, Siberian division
of Russian
  Academy of Sciences, Novosibirsk, Russia}

\date{\today} \maketitle

\begin{abstract}
Dense ($n=4 \times 10^{11}$ cm$^{-2}$) arrays of Ge quantum dots
in Si host were studied using attenuation of surface acoustic
waves (SAWs) propagating along the surface of a piezoelectric
crystal located near the sample. The SAW magneto-attenuation
coefficient,   $\Delta\Gamma=\Gamma(\omega, H)-\Gamma(\omega, 0)$,
and change of velocity of SAW, $\Delta V /V=(V(H)-V(0)) / V(0)$,
were measured in the temperature interval $T$ = 1.5-4.2 K as a
function of magnetic field $H$ up to 6 T for the waves in the
frequency range $f$ = 30-300 MHz. Basing on the dependences of
$\Delta\Gamma$ on $H$, $T$ and $\omega$, as well as on its sign,
we believe that the AC conduction mechanism is a combination of
diffusion at the mobility edge with hopping between localized
states at the Fermi level. The measured magnetic field dependence
of the SAW attenuation is discussed basing on existing theoretical
concepts.

\end{abstract}

\narrowtext

\section{Introduction}
\label{Introduction}

We study Si samples with high-density ($n=4 \times 10^{11}$
cm$^{-2}$) arrays of Ge quantum dots (QD). According to Ref.
\cite{1}, the low-temperature DC conductance of such samples is
due to variable range hopping between different QDs. In such dense
systems the long range Coulomb interaction is very important,
mainly be-cause its influence on the electron density of states
\cite{2}. Both the density of states and the DC conductance are
sensitive to decay length of the electron states localized on the
QD. Hence, it is tempting to find the electron localization length
by an independent method. To fulfill this task we study
attenuation of surface acoustic waves (SAWs) propagating near the
QD array as a function of external magnetic field. SAW attenuation
allows one to determine the AC conductance of the array,  $\sigma
(\omega)$, which depends on the localization length. The external
magnetic field shrinks the electron wave function, and the
consequent shortening of the localization length results in a
decrease in the SAW attenuation.

We are aware only of one experiment \cite{3}, in which SAW
attenuation in a GaAs/AlGaAs array of relatively large (250-500
nm) QDs was studied. In these samples QDs were fabricated using
holographic lithography with subsequent ion etching. The authors
interpreted their results assuming relaxational absorption within
individual QDs \cite{4}. One can expect this mechanism to be very
weakly dependent on magnetic field, so that the magnetic field
dependence of the attenuation would not manifest itself in
relatively weak magnetic fields. Consequently a dependence of the
SAW attenuation on magnetic field at weak fields can be mostly
ascribed to inter-dot transitions.

\section{Experimental Results}

\label{Experiment}

We have measured variation of the attenuation coefficient,
$\Delta\Gamma=\Gamma(\omega, H)-\Gamma(\omega, 0)$ and change of
velocity of SAW $\Delta V /V=(V(H)-V(0)) / V(0)$ - in B-doped "Ge
in Si" dense ($n=4 \times 10^{11}$ cm$^{-2}$) QD arrays in
magnetic field $H \leq$ 6 T.

\begin{figure}[h]
\centerline{\psfig{figure=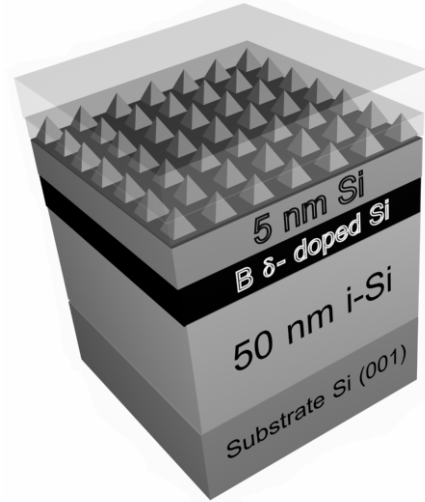,width=6cm,clip=} }
 \caption{Scheme of the quantum dot array. \label{Sample}}
\end{figure}

The samples were grown by MBE method on the (001) Si-substrate.
Firstly, a 50 nm buffer layer of intrinsic Si-doped with B with
$N=2.4 \times 10^{12}$ cm$^{-2}$ (6 holes per QD) was grown. Then
a 5 nm undoped Si layer was grown, on top of which 8 Ge monolayers
were placed. This structure was covered by a 30 nm i-Si layer. The
self-organized Ge QDs had pyramidal shape with height of 15 $\AA$
and square 100 $\times$ 100 $\AA^2$ base Fig.~\ref{Sample}.

\begin{figure}[h]
\centerline{\psfig{figure=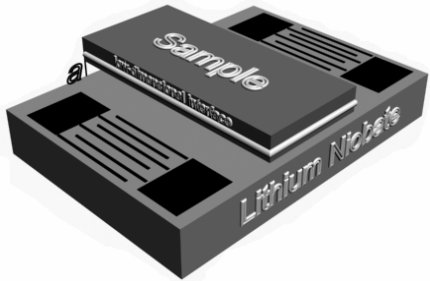,width=6cm,clip=}}
\caption{Scheme of the acoustoelectric device. The electric field
of a surface acoustic wave propagating on the surface of a
piezoelectric substrate acts on a low-dimensional electron system
"embedded" into the sample close to its surface. This "hybrid"
geometry allows applying a sliding electrostatic potential to the
electron/hole system in non-piezoelectric materials \label{SetUp}}
\end{figure}

In our experiments the sample was pressed to a piezoelectric
LiNbO$_3$ crystal by a spring, and SAW propagated along the
crystal surface Fig.~\ref{SetUp}. In such geometry the SAW is
coupled to the holes only by electric fields, and a direct
mechanical coupling turns out to be not very important \cite{5,
6}. The SAW frequency was in the range 30-300 MHz, and the input
intensity varied between 3$\times$10$^{-6}$ and 3$\times$10$^{-3}$
W/cm. Since $\Delta\Gamma$ is small even at $H$=6 T (about few
$\%$), two methods were used to determine the SAW attenuation -
direct measurement of the SAW amplitude $U$ at the receiver and
comparison of this amplitude with the amplitude $U_0$ of another
signal, which passed through the receiver's amplifier avoiding the
sample. In the second case the result is just the ratio $U_0 / U$.

DC conductance was measured in this sample in the temperature
interval 4.2-25 K without magnetic field. In this temperature
interval it obeys the law  $\sigma^{DC}$($\Omega^{-1}$) = $7.8
\times 10^{-5} exp[-(282/T)^{0.5}]$. This dependence is compatible
with the variable range hopping in the Coulomb gap regime
(Shklovskii-Efros mechanism \cite{7}). Having in mind the values
of the DC conductance, we expect that in the relevant frequency
range the AC conductance of the sample is $< 10^{-7}$
$\Omega^{-1}$. At such values of conductance the screening of SAW
electric field by the layer of QDs is not important, and the
attenuation is just proportional to real part of the complex
conductance,  $\Gamma (\omega) \propto Re \sigma_{xx} (\omega)$.

\begin{figure}[h]
\centerline{\psfig{figure=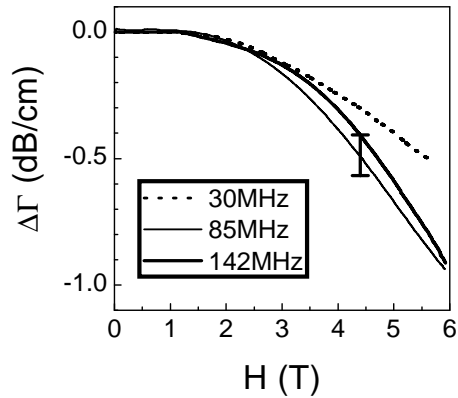,width=6cm,clip=}}
\caption{Dependence of  $\Delta\Gamma=\Gamma(\omega,
H)-\Gamma(\omega, 0)$, $T$=4.2 K \label{GamMF}}
\end{figure}

The measured magneto-attenuation $\Delta\Gamma$ as a function of
$H$ for different SAW frequencies is shown in Fig.~\ref{GamMF}.
One can see that the attenuation {\it decreases} with magnetic
field, $\Delta\Gamma<$0. The absolute value of $\Delta\Gamma$ is
proportional to $H^2$ (Fig.~\ref{GamMF2}).

To make the method useful for quantitative studies of the material
one has to extract  $\Delta \sigma (\omega)$ and find its
frequency and temperature dependences. For this purpose we employ
the model of Ref. \cite{8}, which presents the system as a
combination of semi-infinite substrate and sample divided by a
vacuum clearance with thickness $a$. For $\sigma < 10^{-7}
\Omega^{-1}$:

\begin{figure}[h]
\centerline{\psfig{figure=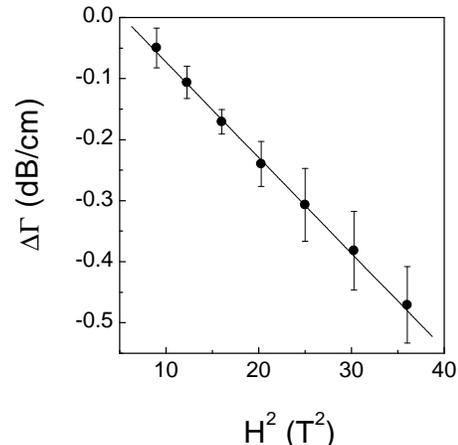,width=6cm,clip=}}
\caption{Illustrative dependence of $\Delta\Gamma$ on $H^2$.
$f$=30 MHz, $T$=4.2K \label{GamMF2}}
\end{figure}

\begin{eqnarray}
\Delta \Gamma=\frac{\Delta\sigma A(q)q e^{-2qa}}{2[(\varepsilon_0+
\varepsilon_1) (\varepsilon_s + \varepsilon_0)-
(\varepsilon_1-\varepsilon_0)(\varepsilon_s -
\varepsilon_0)e^{-2qa}]^2} ,  
\end{eqnarray}
\begin{eqnarray}
A(q)= 8.68  \frac{K^2}{2} 8 (\varepsilon_0+
\varepsilon_1)\varepsilon_0^2 \varepsilon_s , \nonumber
\end{eqnarray}

where $q$ is the SAW wave vector, $K^2$ is the electromechanical
coupling constant of the substrate (LiNbO$_3$),
$\varepsilon_1$=51, $\varepsilon_0$=1, $\varepsilon_s$=12 are the
dielectric constants of LiNbO$_3$, vacuum and the sample,
respectively. The clearance $a=3\times 10^{-5}$ cm was determined
using the well-known thin-film light interference method.

\begin{figure}[h]
\centerline{\psfig{figure=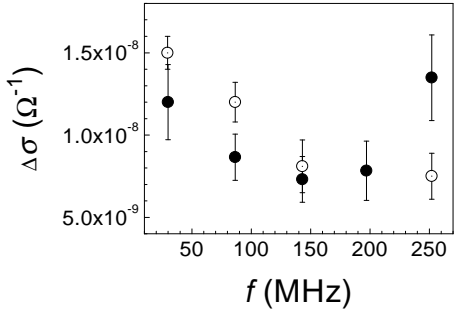,width=6cm,clip=}}
\caption{Frequency dependence of $\Delta\sigma$, $T$=4.2 K, $H$=6
T measured on two samples at different settings (with clearance
being $a=2$ and $3 \times 10^{-5}$ cm). \label{SigFreq}}
\end{figure}

The dependence  $\Delta\sigma(\omega)$ at $H$=6 T and $T$=4.2 K,
obtained using the aforementioned procedure, is shown in
Fig.~\ref{SigFreq}. The error was 30-40$\%$. One can conclude that
there is no pronounced frequency dependence of $\Delta\sigma$ in
the frequency range 30-270 MHz.

\begin{figure}[h]
\centerline{\psfig{figure=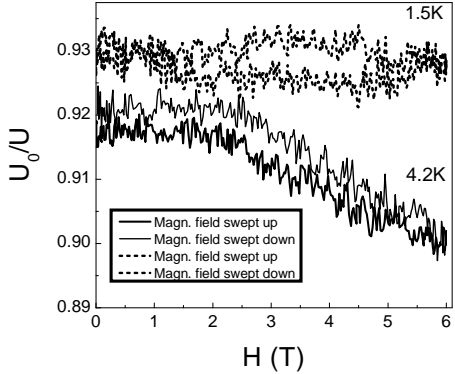,width=6cm,clip=}}
\caption{Dependence of measured in experiments value of $U_0 / U$
on magnetic field at $T$=1.5 K and 4.2K, $f$=87MHz. \label{Umf}}
\end{figure}

To get an idea about temperature dependence of the
magneto-attenuation we compare the ratio $U_0 / U$ at frequency 87
MHz for temperatures 4.2 K and 1.5 K (Fig.~\ref{Umf}). Decrease of
this ratio with increase of $H$ means decreasing of the SAW
attenuation. At $T$=1.5 K the SAW attenuation is practically
magnetic field-independent. Different curves correspond to
increasing and decreasing branches of the magnetic field cycle in
the range 0-6 T.

As regards the measurements of the velocity of SAW in magnetic
field this effect was small to be measured (relative change of
velocity me can measure with accurate within $5 \times 10^{-5}$).

\section{Discussion}

We have observed negative SAW magneto-attenuation, which is
proportional to $H^2$ in magnetic fields up to 6 T and in the
temperature range 1.5-4.2 K. The real part of magnetoconductance,
$\Delta\sigma$, obtained from the raw data using the three-layer
model \cite{8}, is also negative and proportional to $H^2$. It
increases with temperature and almost frequency-independent.

Simple estimates show that the observed behavior cannot be allowed
for taking into account only intra-dot transitions. Indeed, the
inter-level distances in the small dots significantly exceed both
$\hbar\omega$ and $kT$. Thus we are left with two mechanisms: (i)
tunneling or thermally-activated hopping between the states
localized in different dots forming pairs; (ii) activation to the
mobility edge, as in the case of the Coulomb-gap-mediated DC
conductance.

In the first case, the theory (see Ref. \cite{9} for a review)
predicts a crossover from close-to-linear in frequency and
temperature-independent to frequency-independent, but
temperature-dependent $\Delta\sigma$. This crossover takes place
at $\omega\tau \approx1$, where $\tau(T)$ is the typical
relaxation time for the inter-dot transitions. An-other feature of
the inter-dot hopping is big ratio between imaginary and real
parts of the complex conductance, Im$\sigma(H,\omega)$/Re
$\sigma(H,\omega)$$\gg$1. The imaginary part of the
magnetoconductance can be, in principle, extracted from the
measured SAW velocity V. However, the variation of the velocity in
magnetic field is too small to be measured at present time.

In the second case the ratio Im$\sigma(H,\omega)$/Re
$\sigma(H,\omega)$$\ll$1, and the AC magnetoconductance should be
close to the DC one. This behavior could be the case at low enough
frequencies \cite{10}, when the distance between the relevant
localized states, $r_\omega$ , is comparable with the hopping
length, $r_T$, for the DC VRH conductivity. Crude estimations
show, that in our experiment $r_\omega$ has indeed the same order
of magnitude as $r_T$. Consequently, it seems to be difficult to
discriminate between the processes at the mobility edge and the
processes involving localized states.

\bigskip
\section{Conclusions} \label{Conclusion}

For the first time, AC magnetoconductance in a Ge-in-Si QD array
is measured using the SAW technique. Mechanism of AC conductivity
is probably a combination of diffusion at the mobility edge with
hopping between localized states at the Fermi level.

\acknowledgments The work is supported by RFBR (Grants No.
01-02-17891 and 03-03-16536), MinNauki, Presidium RAN grants,
NATO-CLG.979355 grants.


\widetext


\begin{thebibliography}{99}

\bibitem{1} A. I. Yakimov, A. V. Dvurechenskii, R. Boucher,
A. V. Dvurechenski, J. Phys.: Condens. Matter \textbf{11}, 9715
(1999).

\bibitem{2} I. Yakimov, A. V. Dvurechenski, V. V. Kirienko,
Yu. I. Yakovlev, Phys. Rev. B \textbf{61}, 10868 (2000).

\bibitem{3} G. R. Nash, S. J. Bending, M. Boero,
M. Riek, K. Eberl, Phys.Rev. B \textbf{59}, 7649 (1999).

\bibitem{4} A. Knabchen, O. Entin Wohlman,
Y. M. Galperin, Y. B. Levinson, Europ. Lett. \textbf{39}, 419
(1997).

\bibitem{5} I. L. Drichko,  A. M. Diakonov, A. M. Kreshchuk,
T. A. Polyanskaya, I. G. Savel'ev, I. Yu. Smirnov, A. V. Suslov,
Semiconductors \textbf{31}, 384 (1997) [FTP \textbf{31}, 451
(1997)].

\bibitem{6} I. L. Drichko, A. M. Diakonov, I. Yu. Smirnov,
Y. M. Galperin, and A. I. Toropov, Phys. Rev. B \textbf{62}, 7470
(2000).

\bibitem{7} A. L. Efros, B. I. Shklovskii,
in: "Electron-electron interactions in Disordered Systems",
ed. by A. L. Efros and M. Pollak, Elsevier, B.V. p.409 (1985).

\bibitem{8} V. D. Kagan, Semiconductors \textbf{31}, 470 (1997).

\bibitem{9} Y. M. Galperin, V. L. Gurevich and
D. A. Parshin, in "Hopping Transport in Solids",
ed. By M. Pollak and B. I. Shklovskii, North-Holland, p. 81 (1991).


\bibitem{10} A. L. Efros, ZETF \textbf{89}, 1834 (1985) [Sov. Phys. JETP
\textbf{89}, 1057 (1985)].



\end{thebibliography}
\end{document}